\documentclass[aps,pre,twocolumn,showpacs,preprintnumbers,amsmath,amssymb]{revtex4}
\usepackage{epsfig}

\begin{document}
\title{Kauffman Boolean model in undirected scale free networks}
\author{Piotr Fronczak, Agata Fronczak and Janusz A. Ho\l yst}
\affiliation{Faculty of Physics and Center of Excellence for
Complex Systems Research, Warsaw University of Technology,
Koszykowa 75, PL-00-662 Warsaw, Poland}
\date{\today}

\begin{abstract}
We investigate analytically and numerically the critical line in
undirected random Boolean networks with arbitrary degree distributions,
including scale-free topology of connections $P(k)\sim
k^{-\gamma}$. We show that in infinite scale-free
networks the transition between frozen and chaotic phase occurs
for $3<\gamma < 3.5$. The observation is interesting for two
reasons. First, since most of critical phenomena in scale-free
networks reveal their non-trivial character for $\gamma<3$, the
position of the critical line in Kauffman model seems to be an
important exception from the rule. Second, since gene regulatory
networks are characterized by scale-free topology with $\gamma<3$,
the observation that in finite-size networks the mentioned
transition moves towards smaller $\gamma$ is an argument for
Kauffman model as a good starting point to model real systems. We
also explain that the unattainability of the critical line in
numerical simulations of classical random graphs is due to
percolation phenomena.
\end{abstract} \pacs{89.75.Hc, 89.75.-k, 64.60.Cn, 05.45.-a} \maketitle

Almost 40 years ago Stuart Kauffman proposed random Boolean
networks (RBNs) for modelling gene regulatory networks
\cite{kauffman1969}. Since then, beside its original purpose, the
model and its modifications have been applied to many different
phenomena like cell differentiation \cite{huang2000}, immune
response \cite{kauffman1989}, evolution \cite{bornholdt1998}, opinion formation \cite{lambiotte},
neural networks \cite{wang1990}, and even quantum gravity problems
\cite{baillie1994}.

The original RBNs were represented by a set of $N$ elements,
$\sum_t=\{\sigma_1(t), \sigma_2(t),...,\sigma_N(t)\}$, each element 
$\sigma_i$ having two possible states: active ($1$), or inactive ($0$). The
value of $\sigma_i$ was controlled by $k$ other elements of the
network, i.e. $\sigma_i(t+1)=f_i(\sigma_{i_1}(t),
\sigma_{i_2}(t),...,\sigma_{i_k}(t))$, where $k$ was a fixed
parameter. The functions $f_i$ were selected so that they have returned
values $1$ and $0$ with probabilities respectively equal to $p$
and $1-p$. The parameters $k$ and $p$ have determined the dynamics
of the system (Kauffman network), and it has been shown that for a
given probability $p$, there exists the critical number of inputs
\cite{derrida1986}
\begin{equation}\label{eq_kauffman}
k_c=\frac{1}{2p(1-p)},
\end{equation}
below which all perturbations in the initial state of the system
die out ({\it frozen phase}), and above which a small perturbation
in the initial state of the system may propagate across the entire
network ({\it chaotic phase}).

In fact, the behavior of Kauffman model in the vicinity of the
critical line $k_c(p)$ has become a major
concern of scientists interested in gene regulatory networks. The
main reason for this was the conjecture that living organisms
operate in a region between order and complete randomness or chaos
(the so-called {\it edge of chaos}) where both complexity and rate
of evolution are maximized \cite{kauffman1990,sole2001,stauffer1994}. The analogous behavior has been noticed in Kauffman networks, which in the interesting region described by eq. (\ref{eq_kauffman}) show stability, homeostatis, and the ability to cope with minor modifications when mutated. The networks are stable as well as flexible in this region. 

Recently, when data from real networks have become available \cite{albert2002,barabasi2002}, a quantitative comparison of the
{\it edge of chaos} in these datasets and RBN models has brought an
encouraging and promising message that even such simple model may
quite well mimic characteristics of real systems. 

Since, however, one has noticed that real genetic networks exhibit a wide range of
connectivities, the recent modifications of the standard RBN take into
consideration a distribution of nodes' degrees $P(k)$. It has been
shown that if the random topology of the directed network is homogeneous
(i.e. all elements of the network are statistically equivalent), then the network topology can be meaningfully
characterized by the average in-degree $\langle k\rangle$, and the transition between frozen and chaotic phase occurs for \cite{sole1997}:
\begin{equation}\label{eq_sole}
\langle k\rangle_c=\frac{1}{2p(1-p)}.
\end{equation}

On the other hand, if the network topology is characterized by a wide
heterogeneity in the connectivity of elements, then it is useless to
characterize the network by the average in-degree, and instead of $\langle k\rangle$ another parameter must be used. In the case of power-law in-degree distribution 
$P(k)=[\zeta(\gamma)k^\gamma]^{-1}$, where $\zeta(\gamma)$ is the zeta function, the characteristic exponent $\gamma$ is the relevant parameter. It has been shown that the critical line $\gamma_c(p)$ in RBN model defined on scale-free networks is given by \cite{aldana2003physd}:
\begin{equation}\label{eq_aldana}
\frac{\zeta(\gamma_c-1)}{\zeta(\gamma_c)}=\frac{1}{2p(1-p)}.
\end{equation}
Since $2<\gamma_c(p)<2.5$, based on the result (\ref{eq_aldana}) it was claimed \cite{aldana2003pnas} that the abundance of scale-free networks with $2<\gamma<3$ in nature and society can be attributed to the presence of both phases, frozen and chaotic, only is such networks.

Recently, several authors \cite{lee,boguna} have provided a general formula for the edge of chaos in directed networks characterized by the joint degree distribution $P(k,q)$
\begin{equation}\label{lee_eq}
\frac{\langle kq\rangle}{\langle q\rangle}=\frac{1}{2p(1-p)},
\end{equation}
where $k$ and $q$ correspond to in- and out-degrees of the same node, respectively. The formula (\ref{lee_eq}) shows that the position of the critical line depends on the correlations between $k$ and $q$ in such networks. It is also easy to show that the previous results (\ref{eq_kauffman})-(\ref{eq_aldana}) immediately follow from (\ref{lee_eq}) if one assumes the lack of correlations $P(k,q)=P_{in}(k)P_{out}(q)$. 


In this paper, we derive general relation describing position of the critical line in undirected RBNs with arbitrary distribution of connections $P(k)$. The specific cases, including homogeneous as well as strongly heterogeneous (i.e. scale-free) random network topologies are discussed. We also generalize our derivations to the case when the
scale-free network topology is characterized not only by the
exponent $\gamma$ but also by the minimal node degree $k_{min}=m$,
which controls the density of connections. We show that for
$\gamma\rightarrow\infty$ the parameter $m$ corresponds to the original
parameter $k$ used in the standard Kauffman model defined on regular random graphs, in which the number of connections is the same for all elements. 

In order to find the position of the critical line in RBN one has to examine the sensitivity of its dynamics with regard to the initial conditions. In numerical studies such a sensitivity can be analyzed quite simply. One has to start with two initial states $\sum_0=\{\sigma_1(0), \sigma_2(0),...,\sigma_N(0)\}$ and
$\widetilde{\sum}_0=\{\widetilde{\sigma}_1(0),
\widetilde{\sigma}_2(0),...,\widetilde{\sigma}_N(0)\}$, which are identical except for a small number of elements, and observe how the differences between both configurations $\sum_t$ and $\widetilde{\sum}_t$ change in time. If a system is robust then the studied configurations lead to similar long-time behavior, otherwise the differences develop in time. A suitable measure for the distance between the configurations is the overlap $x(t)$ defined as
\begin{equation}
x(t)=1-\frac{1}{N}\sum_{i=1}^N|\sigma_i(t)-\widetilde{\sigma}_i(t)|.
\end{equation}
Note, that in the limit $N\rightarrow\infty$, the overlap becomes the probability for two arbitrary but corresponding elements, $\sigma_i(t)$ and $\widetilde{\sigma_i}(t)$, to be equal. Moreover, the stationary long-time limit of the overlap $x=\lim_{t\rightarrow\infty}x(t)$ can be treated as the order parameter of the system. If $x=1$ then the system is insensitive to initial perturbations (frozen phase), while for $x<1$, the initial perturbations propagate across the entire network (chaotic phase).

In the following, we will partially reproduce the annealed computation (for the first time carried out by Derrida and Pomeau \cite{derrida1986}), and generalize it to the case of undirected random graphs with arbitrary degree distribution. The case of directed networks has been studied by Aldana \cite{aldana2003physd}, and also by Lee and Rieger \cite{lee}. 

Thus, having in mind that $x(t)$ corresponds to the probability that a given element $i$ possesses the same value in both configurations, $\sigma_i(t)=\widetilde{\sigma}_i(t)$, two different situations have to be considered. If all the $k_i$ inputs of $\sigma_i(t)$ are equal to respective inputs of $\widetilde{\sigma}_i(t)$, which occurs with probability $[x(t)]^{k_i}$, then one has $\sigma_i(t+1)=\widetilde{\sigma}_i(t+1)$.
On the other hand, if at least one of the $k_i$ inputs of
$\sigma_i(t)$ differs from its counterpart in $\widetilde{\sum}_t$,
which occurs with probability $1-[x(t)]^{k_i}$, then
$\sigma_i(t+1)=\widetilde{\sigma}_i(t+1)$ only if
$f_i(\sigma_{i_1}(t),...,\sigma_{i_{k_i}}(t))=f_i(\widetilde{\sigma}_{i_1}(t),...,\widetilde{\sigma}_{i_{k_i}}(t))$
regardless of the values of the inputs in each configuration.
Probability of such an event is $p^2+(1-p)^2$.
Taking all the above together one finds that the probability $x(t+1)$ that
$\sigma_i(t+1)=\widetilde{\sigma}_i(t+1)$ is given by
\begin{eqnarray}\label{dlugie}
x(t+1)&=&\sum_{k_i=1}^{\infty}\{[x(t)]^{k_i}\cdot 1+\nonumber \\&&(1-[x(t)]^{k_i})\cdot (p^2+(1-p)^2)\}Q(k_i),
\end{eqnarray}
where $Q(k_i)$ represents probability that an arbitrary link leads to the node $i$ of degree $k_i$. In uncorrelated networks $Q(k)$ corresponds to the degree distribution of the nearest neighbors
\begin{equation}
Q(k)=\frac{k}{\langle k\rangle}P(k).
\end{equation}

The equation (\ref{dlugie}) can be understood as a map $x(t+1)=M(x(t))$, where
\begin{equation}\label{mapa}
M(x)\equiv 1-2p(1-p)\{1-\sum_{k=1}^\infty x^{k} Q(k)\}
\end{equation}
It can be shown that the change of the stability of the fixed
point of the map $x=M(x)$, which occurs when
\begin{equation}\label{limes}
lim_{x\rightarrow 1^-}\frac{dM(x)}{dx}=1,
\end{equation}
determines the phase transition between the ordered and chaotic
regimes (c.f. \cite{aldana2003physd}). Substituting (\ref{mapa}) into (\ref{limes}) one gets the
condition for the phase transition:
\begin{equation}\label{nasze}
\frac{\langle k^2 \rangle}{\langle k \rangle}=\frac{1}{2p(1-p)}.
\end{equation}
In the following we will analyze the equation (\ref{nasze}) in classical random graphs and in scale-free networks where the second moment $\langle k^2 \rangle$ becomes important (it diverges for $\gamma<3$).

Since in classical random graphs $\langle k^2 \rangle=\langle k \rangle^2+\langle k \rangle$, the eq. (\ref{nasze}) simplifies:
\begin{equation}\label{eq_ER}
\langle k\rangle_c=\frac{1}{2p(1-p)}-1.
\end{equation}

\begin{figure}
 \centerline{\epsfig{file=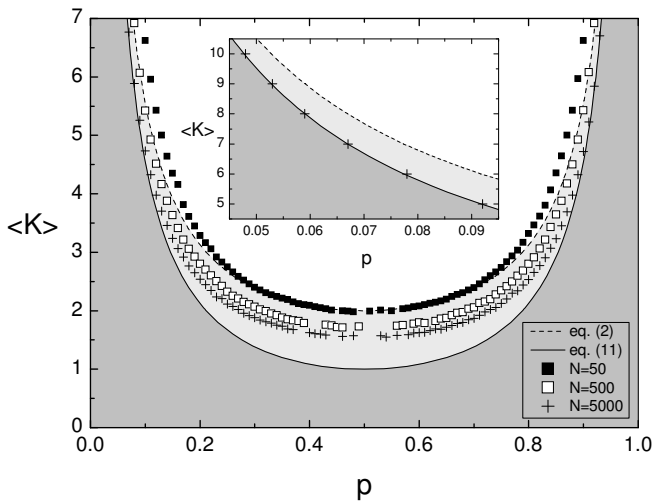,width=\columnwidth}}
    \caption{Phase diagram of Kauffman model defined on classical random graphs. Frozen phase resulting from eq. (\ref{eq_ER}) is marked by the dark gray area. Light gray area shows the difference between directed and undirected networks.}
        \label{fig1}
\end{figure}
Comparing the formula (\ref{eq_ER}) with (\ref{eq_sole}) one can see that the critical curve in undirected networks has been shifted by $1$ in comparison with the directed case. The figure \ref{fig1} presents both equations as well as numerical simulations of undirected networks of three different sizes  ($N=50$, $N=500$, and $N=5000$). While in the limit of large $\langle k \rangle$ the results, especially for large $N$, agree very well with the eq. (\ref{eq_ER}) (see inset), for $\langle k \rangle \rightarrow 1$ (i.e. $p\rightarrow 0.5$) they differ significantly. The discrepancy results from the fact that $\langle k \rangle =1$ corresponds to the percolation threshold in these networks. Because the size of the largest component near $\langle k \rangle =1$ is significantly smaller than the network size (the network is divided into several not connected components), any perturbation cannot propagate across the entire system, and the frozen phase is easier achieved. It means that it is impossible to verify eq. (\ref{eq_ER}) in this range. The closer percolation threshold we are the smaller networks (separated pieces of the whole network) we analyze. One can also show that if one introduces assortativity (i.e. positive degree-degree correlations) to the network the attainable critical connectivity can be significantly shifted towards $\langle k \rangle_c=1$. It happens because the percolation transition occurs for lower values of $\langle k \rangle$ in assortative networks \cite{newman2002,noh2007}. Unfortunately, due to the introduced correlations, analytical treatment is much more difficult in such a case.

\begin{figure}[t]
 \centerline{\epsfig{file=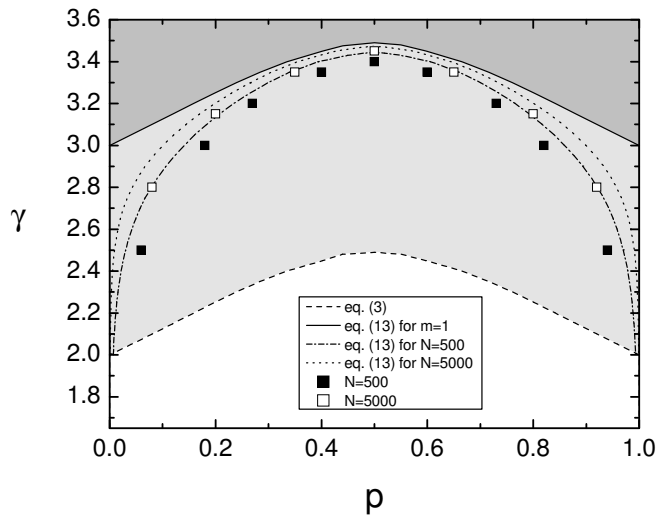,width=\columnwidth}}
    \caption{Phase diagram of scale-free networks with $m=1$. Frozen phase resulting from eq. (\ref{eq_SF}) is marked by dark gray area. Light gray area shows the difference between directed and undirected networks. Points represent results of numerical simulations, while the two intermediate lines are solutions of eq. (\ref{eq_SF}) modified for finite networks (dot-dashed line for $N=500$, and dotted line for $N=5000$).}
        \label{fig2}
\end{figure}

Now, let us analyze scale-free networks with the degree distribution given by power law
\begin{equation}\label{SF}
P(k)=[\zeta(\gamma,m)k^\gamma]^{-1},
\end{equation}
where $\zeta(\gamma,m)=\sum_{k=m}^\infty k^{-\gamma}$ is the generalized Riemann zeta function (normalization factor), and the parameter $m$ represents the minimal node degree, i.e. it controls density of connections in the considered networks. Now the eq. (\ref{nasze}) takes a form:
\begin{equation}\label{eq_SF}
\frac{\zeta(\gamma-2,m)}{\zeta(\gamma-1,m)}=\frac{1}{2p(1-p)}.
\end{equation}

\begin{figure}[th]
 \centerline{\epsfig{file=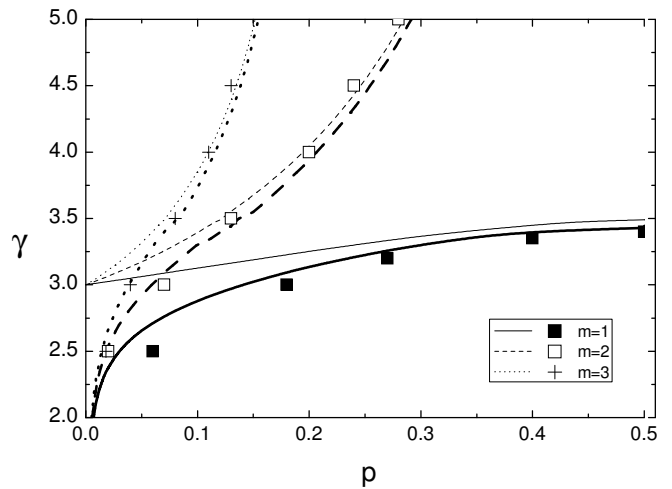,width=\columnwidth}}
    \caption{Critical lines for scale-free RBNs with $m=1$ (solid lines and filled points), $m=2$ (dashed lines and open points) and $m=3$ (dotted lines and crosses). Thick lines are solutions of eq. (\ref{eq_SF}), while the thin lines are solutions of the same equation modified for networks of size $N=5000$. Points correspond to the results of numerical simulations.}
        \label{fig3}
\end{figure}

\begin{figure}[th]
 \centerline{\epsfig{file=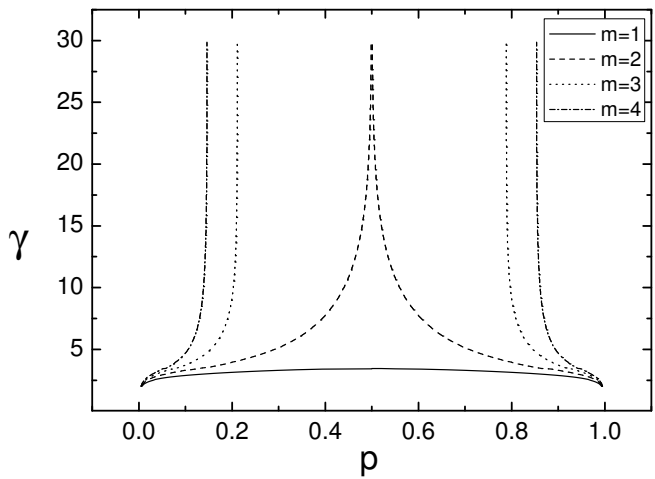,width=\columnwidth}}
    \caption{Phase diagram of scale-free networks with different values of the parameter $m$.}
        \label{fig4}
\end{figure}

In figure \ref{fig2} comparison of transcendental equations (\ref{eq_SF}) (undirected network) for $m=1$ and (\ref{eq_aldana}) (directed network) is presented. Analytical curves taking into account finite size version of the distribution (\ref{SF}) (where zeta functions have been replaced by finite sums), as well as results of the numerical simulations for $N=500$ and $N=5000$ are also shown in the figure. One can see that in undirected case of infinite scale-free networks the transition between frozen and chaotic phase occurs for $3< \gamma < 3.5$. It means that in the studied network the critical line has been shifted in comparison with the directed case by $\Delta\gamma=1$ towards larger values of the exponent $\gamma$. 

The observation is interesting for two
reasons. First, since gene regulatory
networks are characterized by scale-free topology with $\gamma<3$,
the observation that in finite-size networks the mentioned
transition moves towards smaller $\gamma$ is an argument for
Kauffman model as a good starting point to model real systems. 
Second, most of critical phenomena in scale-free
networks reveals its non-trivial character for $\gamma<3$ making these networks interesting for researchers \cite{dorogov}. It happens because the second moment of the degree distribution is size dependent for $\gamma<3$ (it diverges for $N\rightarrow\infty$). For example, in the case of percolation transition, the above causes that it is practically impossible to eliminate the giant connected component in such networks, i.e. they are ultraresilient against random damage or failures \cite{albert2000, cohen2000}. It also implies the lack of epidemic threshold in such networks, i.e. the networks are prone to the spreading and the persistence of infections whatever the epidemic spreading rate is. Finally, in Ising model defined on scale-free networks with $\gamma<3$ the critical temperature is size dependent.   
Taking all the above into consideration the position of the critical line in Kauffman model shows that scale-free networks with $\gamma>3$ may also exhibit interesting properties.

In previous papers \cite{aldana2003physd,aldana2003pnas} it has been stated that the only natural parameter which determines the network topology is the scale-free exponent $\gamma$. In this paper, we introduce the parameter $m$, which does not change the scale-free character of the node degree distribution, but allows us to control the density of connections. For $m=1$ we retrieve the original problem studied in \cite{aldana2003physd,aldana2003pnas}.
In figure \ref{fig3} and \ref{fig4} we present the solutions of the eq. (\ref{eq_SF}) for different values of the parameter $m$. As one can see, for $m>2$ the frozen phase is preserved only for sufficiently small and for sufficiently high values of the parameter $p$. For a wide range of intermediate values of $p$ the frozen phase is unattainable.

\begin{figure}
 \centerline{\epsfig{file=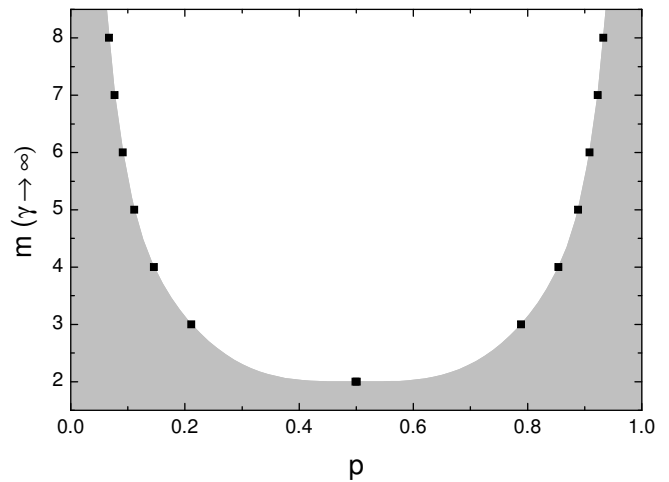,width=\columnwidth}}
    \caption{Phase diagram of scale-free RBNs with $\gamma\rightarrow\infty$. The diagram coincides with the phase diagram of  the standard RBN model. Points for a given $m$ show the width of the chaotic phase taken from the fig. \ref{fig4} for $\gamma=30$.}
        \label{fig5}
\end{figure}

It is worth noting that in the limit $\gamma\rightarrow\infty$, the scale-free distribution (\ref{SF}) transforms into the Dirac delta function $\delta(k-m)$ (then $\langle k^2\rangle=\langle k\rangle^2$ and eq. (\ref{nasze}) simplifies to eq. (\ref{eq_kauffman})). It means that in this limit the scale-free RBN model transforms to the standard RBN model, where all elements have the same node degree. In fig. \ref{fig4} one can see that for $\gamma\rightarrow\infty$ and $m=2$ the width of the chaotic phase shrinks to zero. In figure \ref{fig5} we show this width for different values of the parameter $m$. In this figure one can easily recognize the phase diagram of the standard RBN model, in which for $p=0.5$ the critical value of the node degree equals to $m=k_c=2$. 

In summary, we have investigated analytically and numerically the critical line in undirected random Boolean networks with arbitrary degree distribution including homogeneous and scale-free topology of connections. 
We have shown that in infinite scale-free networks the transition between frozen and chaotic phase occurs for $3<\gamma< 3.5$, i.e. position of the critical line is shifted by $\Delta\gamma=1$ towards larger values of the exponent $\gamma$ in comparison with the directed case. The observation is interesting for two
reasons. First, since most of critical phenomena in scale-free
networks reveals its non-trivial character for $\gamma<3$, the
position of critical line in Kauffman model seems to be an
important exception from the rule. Second, since gene regulatory
networks are characterized by scale-free topology with $\gamma<3$,
the observation that in finite-size networks the mentioned
transition moves towards smaller $\gamma$ is an argument for
Kauffman model as a good starting point to model real systems. We
also explain that the unattainability of the critical line in
numerical simulations of classical random graphs is due to
percolation phenomena.

The work was funded in part by the European Commission Project
CREEN FP6-2003-NEST-Path-012864 (P.F.), the State Committee for Scientific Research in Poland under Grant 1P03B04727 (A.F.), and by the
Ministry of Education and Science in Poland under Grant
134/E-365/6.PR UE/DIE 239/2005-2007 (J.A.H.).

\end{document}